\documentclass[amsbook,12pt]{article}
\usepackage{epsfig, amsfonts,amssymb, amsmath}
\usepackage[usenames]{xcolor} 
\usepackage{picinpar}
\usepackage{float}
\usepackage{graphicx}
\graphicspath{ {./}, {./figures/} }
\usepackage{hyperref}
\usepackage{hyperref}
\newcommand{\nl}{\mbox{}\\}

\begin{document}
\definecolor{OliveGreen1}{rgb}{0,0.6,0}
\definecolor{OliveGreen2}{cmyk}{0.64,0,0.95,0.40}
\definecolor{MoonstoneBlue}{rgb}{0.45, 0.66,0.76}
\definecolor{NavyBlue}{rgb}{0.00, 0.35, 0.50}
%
%
%
\mbox{} \vspace{+0.100cm} \\
\hypersetup{
colorlinks=true,
linkcolor=blue,
filecolor=magenta,
urlcolor=blue,
}
\urlstyle{same}
%
%
\mbox{} \hspace{-0.750cm} 
\mbox{\Large \bf Herd immunity for Covid-19 in homogenous populations} \\
\begin{center}
\mbox{} \vspace{-0.500cm} \\
{\sc Jana\'\i na P\!\;\!. Zingano, Paulo R. Zingano,} \\
\mbox{} \vspace{-0.575cm} \\
{\footnotesize Institute of Mathematics and Statistics} \\
\mbox{} \vspace{-0.700cm} \\
{\footnotesize Universidade Federal do Rio Grande do Sul} \\
\mbox{} \vspace{-0.700cm} \\
{\footnotesize Porto Alegre, RS 91509-900, Brazil} \\
\mbox{} \vspace{-0.150cm} \\
{\sc Alessandra M. Silva} \\
\mbox{} \vspace{-0.575cm} \\
{\footnotesize Companhia de Planejamento do Distrito Federal} \\
\mbox{} \vspace{-0.700cm} \\
{\footnotesize Governo de Bras\'\i lia} \\
\mbox{} \vspace{-0.700cm} \\
{\footnotesize Bras\'\i lia, DF 70620-080, Brazil} \\
\mbox{} \vspace{-0.350cm} \\
{\sc and} \\
\mbox{} \vspace{-0.350cm} \\
{\sc Carolina P\!\:\!. Zingano} \\
\mbox{} \vspace{-0.600cm} \\
{\footnotesize School of Medicine} \\
\mbox{} \vspace{-0.700cm} \\
{\footnotesize Universidade Federal do Rio Grande do Sul} \\
\mbox{} \vspace{-0.700cm} \\
{\footnotesize Porto Alegre, RS 90035-003, Brazil} \\
\nl
\mbox{} \vspace{-0.250cm} \\
{\bf Abstract} \\
\begin{minipage}[t]{12.750cm} 
{\small
\mbox{} \hspace{+0.250cm}
In this note we estimate herd immunity levels
for the Covid-19 epidemic 
based upon a standard SEIR system that models
the disease dynamics in homogeneous populations.
The results obtained are indicative of values 
between 80\:\!\% and 90\:\!\%
for unprotected, fully susceptible populations.
Basic protective measures
like hand hygiene and mask wearing
may be effective to lower herd immunity estimates
down to values between 50\:\!\% and 60\:\!\%.
%
%
}
\end{minipage}
\end{center}
\nl
\mbox{} \vspace{-0.100cm} \\
\mbox{} \hspace{+0.250cm}
\begin{minipage}[t]{14.000cm}
{\bf Key words:}
{\small
Covid-19 outbreak, SARS-Cov-2 coronavirus,
reproduction numbers, \\
\mbox{} \hfill
SEIR models,
homogeneous populations,
herd immunity against Covid-19 \\
}
\end{minipage}
\nl
\mbox{} \vspace{-0.250cm} \\
\mbox{} \hspace{+0.250cm}
\begin{minipage}[t]{14.000cm}
{\bf AMS Subject Classification:}
{\small
92-08, 92-10 (primary), 92-04 (secondary) \\
}
\end{minipage}
\nl
\mbox{} \vspace{-0.550cm} \\
\nl
\mbox{} \hspace{+0.250cm}
\begin{minipage}[t]{14.000cm}
{\bf Matlab code:}
{\small
%
%
\end{minipage}
\nl
\mbox{} \vspace{-0.150cm} \\
\nl
{\small 
\mbox{} \hspace{+1.150cm} 
Corresponding author: 
\begin{minipage}[t]{12.500cm}
Paulo R. Zingano 
(ORCID: 0000-0002-5074-9146)\\
E-mail: paulo.zingano@ufrgs.br 
\end{minipage}
}
\newpage
%
%

%
%
%
%
%
\mbox{} \vspace{-0.750cm} \\
\definecolor{BeauBlue}{rgb}{0.74, 0.83, 0.90}
\definecolor{LightBlue}{rgb}{0.94, 0.99, 0.97}
\definecolor{AirForceBlue}{rgb}{0.36, 0.54, 0.66}
\definecolor{NavyBlue}{rgb}{0.0, 0.47, 0.75}
\definecolor{LightBlue2}{rgb}{0.64, 0.87, 0.93}
\definecolor{DarkBlue}{rgb}{0.00, 0.10, 0.50} 
\definecolor{LemonChiffon}{rgb}{1.0, 0.98, 0.8}
\mbox{} \hspace{+3.950cm}
\fcolorbox{red}{LemonChiffon}
{\begin{minipage}[t]{7.000cm}
 \mbox{} \\
 \mbox{} \vspace{-0.350cm} \\
 \mbox{} \;\;\;\;\;{\bf Summary \;\!and\;\! Conclusions} \\
 \mbox{} \vspace{-0.550cm} \\
\end{minipage}
}
\nl
\nl
\mbox{} \vspace{-0.150cm} \\
\nl
%
%
\mbox{} \hspace{-0.475cm} 
\fcolorbox{NavyBlue}{LightBlue}
{%
\mbox{}\;\;\;\;\!
\begin{minipage}[t]{15.000cm}
\mbox{} \\
\mbox{} \vspace{-0.500cm} \\
\textcolor{black} 
{%
{\bf Section 1. Introduction} \\
Simple models, with fewer variables and parameters,
have practical advantages like involving 
less complex
{\small \sc missing data} and {\small \sc parameter determination}
problems. \\
\mbox{} \vspace{-0.750cm} \\
} 
\end{minipage}
\;\;\;\!
}
\nl
\mbox{} \vspace{-0.350cm} \\
%
%
\mbox{} \hspace{-0.475cm} 
\fcolorbox{NavyBlue}{LightBlue}
{%
\mbox{}\;\;\;\;\!
\begin{minipage}[t]{15.000cm}
\mbox{} \\
\mbox{} \vspace{-0.500cm} \\
\textcolor{black} 
{%
{\bf Section 2. Implementing the {SEIR} model} \\
A critical part is the {\small \sc initialization problem},
that is, a satisfactory determination of 
{\small \sc initial values} for all the
variables in the model. 
For the SEIR system considered, \linebreak
$S(t_0)$ and $D(t_0)$ are given
(modulo some errors), but
$ E(t_0), \:\!I(t_0), \:\!R(t_0) $
are missing. \linebreak
They can be obtained at some
later {\small \sc initial time} 
$\;\!t_{\ast} \!\,\!=\:\! t_0 \!\;\!+ p$
(with $p>0$ depending on the initial guess
for the missing data)
by continuously fitting the model
(that is, the parameters $ \beta(t) $ and $r(t)$, in this case) 
to the available data at $\;\! t_0, t_0+1,\!\;\!...,t_0+p $. \linebreak
Computations can then proceed from $\:\! t_{\ast} \!\;\! $ onwards;
previous values for $ E(t), I(t), R(t) $ 
and related quantities (e.g.,\;{\small \sc transmission rates}
and {\small \sc reproduction numbers})
are not generally reliable. 
For the Covid-19 examples examined, we found
$ 10 < \:\!p < 20 $. \\
} 
\mbox{} \vspace{-0.750cm} \\
\end{minipage}
\;\;\;\!
}
\nl
\mbox{} \vspace{-0.350cm} \\
%
%
\mbox{} \hspace{-0.475cm} 
\fcolorbox{NavyBlue}{LightBlue}
{%
\mbox{}\;\;\;\;\!
\begin{minipage}[t]{15.000cm}
\mbox{} \\
\mbox{} \vspace{-0.425cm} \\
\textcolor{black} 
{%
{\bf Section 3. Herd Immunity for Covid-19} \\
For Covid-19,
we have found higher 
herd immunity levels than previously suggested: 
80\:\!{\small \%} to 90\:\!{\small \%} of 
an (off\;\!-\;\!guard) homogenenous population.
\!\;\!As a general guide, 
basic protective measures 
like hand hygiene and mask wearing
can be effective to reduce \linebreak
herd immunity numbers 
to between 50\:\!{\small \%}
and 70\:\!{\small \%}
of the population,
depending on \linebreak
the transmissibility status
that will hold in this case. 
As transmissibility 
may vary
significantly across different regions, 
local analysis is advised.\\
%
%
} 
\mbox{} \vspace{-0.750cm} \\
\end{minipage}
\;\;\;\!
}
\nl
\mbox{} \vspace{-0.350cm} \\
%
%
\mbox{} \hspace{-0.500cm} 
\fcolorbox{NavyBlue}{LightBlue}
{%
\mbox{}\;\;\;\;\!
\begin{minipage}[t]{15.000cm}
\mbox{} \\
\mbox{} \vspace{-0.475cm} \\
\textcolor{black} 
{%
{\bf Section 4. Closing Remarks} \\
Herd immunity estimates depend on the model used,
data quality
and other factors, some of which hardly predictable.
Results should  always be viewed with caution. \linebreak
Ours indicate
that herd immunity may be close to 90\;\!\mbox{\small \%}
in homogeneous populations,
providing upperbounds for 
more realistic studies taking heterogeneity into account. 
} 
\mbox{} \vspace{-0.750cm} \\
\end{minipage}
\;\;\;\!
}
\newpage
%

%
%
\mbox{} \vspace{-1.250cm} \\
%
%
%
%
%

{\bf 1. Introduction} \\

There is now a well developed 
mathematical literature
providing a rich variety of 
continuous or discrete 
models
and techniques to
investigate the dynamics 
of communicable diseases,
see e.g.\;\cite{Brauer2008, Diekmann2013, %
Hethcote2000, Li2018, Martcheva2015} 
and references therein. 
Among the many \linebreak
important concepts
are {\small \sc reproduction numbers}
and {\small \sc herd immunity},
which are related to 
the disease potential of developing
significant outbreaks.  
{\small \sc Herd immunity} 
is the state
when an infected individual
in a given population 
generates on average less than
one secondary infection
during his infectious period
--- or, in other words,
the effective reproductive number $\!\;\!R_t$ 
is smaller than one
\cite{Brauer2008, Diekmann2013, Fine2011, Randolph2020}.
In this situation, an outbreak will not develop. 
This is achieved when
a proportion $p > \hat{p} \in (\:\!0, 1) $
of the population in question
is immune to the infectious agent,
where $ \hat{p} $ is the 
{\small \sc herd immunity threshold}
\cite{Fine2011, Fontanet2020, Frederiksen2020, %
Omer2020, Randolph2020}.
Immunization may be acquired by overcoming some previous natural
infection or through vaccination. 
%
%
Estimating $ \hat{p} $ 
is a very important problem 
in the study of infectious diseases.
%
%

%
Herd immunity for Covid-19 has
been examined in a number of previous works,
with $ \hat{p} $ estimates
typically lying between 60\:\!{\small \%} and
85\:\!{\small \%}  
for well-mixed, homogeneous populations
\cite{Britton2020, Fontanet2020, %
Frederiksen2020, Kwok2020, Salje2020}.
\!These estimates depend on 
reliable values for the
re\-productive numbers involved
and other assumptions, 
which cannot be guaranteed. 
For example,
in \cite{Kwok2020}
it is estimated
that $ \hat{p} \approx 81\;\!\mbox{\small \%} $
for Spain
using data on the daily number of new
{\small COVID}-19 cases 
and the {\small \sc exponential growth method}
\cite{Bettencourt2008, Diekmann2013, Wallinga2007} to
estimate $ \!\;\!R_t $, 
the effective reproduction number,
along with estimations on
the serial interval 
and some underlying assumptions
\cite{Nishiura2020}. 
Starting 
with the same data, \linebreak
we estimated $R_t$ from 
daily infected population totals
computed by a careful im\-ple\-mentation of
a standard {\small SEIR} model 
(Section~2), 
then obtaining
\mbox{$ 80\:\!\mbox{\small \%} < \hat{p} < 90\:\!\mbox{\small \%} $}
for the Spanish population,
in the absence of protective measures
(Section~3). 
%
%
%

Our approach to estimate 
herd mmunity threshold values
for {\small C}ovid-19 is based on 
carefully implementing some chosen deterministic model 
so as to yield reliable estimates of the  
information needed, 
such as transmission rates and
reproductive ratios. 
%
%
We illustrate the typical procedures
by considering, for simplicity,
the basic {\small SEIR} system
defined by the equations (1.1) below,
where for convenience 
we have ignored effects like 
birth or migration rates,
deaths by natural or other causes, 
and so forth.
This model
divides the entire population in question
into four classes: the {\em susceptible\/} individuals
(class {\small S}),
those {\em exposed} (class~{\small E}, 
formed by infected people who are still inactive, 
that is, not yet transmitting the disease),
the {\em active infected} or {\em infectious\/} individuals (class {\small I})
and the {\em removed\/} ones.
%
The latter class is formed by the people
who have either 
{\em recovered\/} from the disease (class {\small R})
or have {\em died\/} from it
(class {\small D}). \linebreak
In~the simplest setting,
the dynamics among the 
populations 
$ S(t), E(t), I(t), R(t) $
and $ D(t) $
are given by the differential equations \\
\mbox{} \vspace{-0.500cm} \\
\begin{equation}
\tag{1.1}
\left\{\;
\begin{array}{l}
\mbox{$ {\displaystyle
\frac{dS}{dt} }$}
\;=\;
-\,\beta \,
\mbox{$ {\displaystyle
\frac{S(t)}{N} }$} \, I(t), \\
\mbox{} \vspace{-0.250cm} \\
\mbox{$ {\displaystyle
\frac{dE}{dt} }$}
\;=\;
\beta \,
\mbox{$ {\displaystyle
\frac{S(t)}{N} }$} \, I(t)\,-\, \delta \;\!E(t), \\
\mbox{} \vspace{-0.250cm} \\
\mbox{$ {\displaystyle
\frac{dI}{dt} }$}
\;=\;
\delta \;\!E(t) \,-\, (r + \gamma)\;\!I(t), \\
\mbox{} \vspace{-0.250cm} \\
\mbox{$ {\displaystyle
\frac{dR}{dt} }$}
\;=\;
\gamma \, I(t), \\
\mbox{} \vspace{-0.250cm} \\
\mbox{$ {\displaystyle
\frac{dD}{dt} }$}
\;=\;
r \;\! I(t),
\end{array}
\right.
\end{equation}
\nl
\mbox{} \vspace{-0.400cm} \\
see e.g.\;\cite{Brauer2008, Driessche2002, %
Hethcote2000, Martcheva2015}
for a detailed discussion of the various
terms and their meanings.
The parameters $ \beta $
({\small \sc average transmission rate}) and
$ r $ ({\small \sc average lethality rate}
due to the disease)
vary with $t$ (time, here measured in {\small \sc days}),
but $ \delta $ and $ \gamma $
are positive constants given by \\
\mbox{} \vspace{-0.850cm} \\
\begin{equation}
\tag{1.2}
\gamma \;=\;
\frac{1}{\,\mbox{\small $T$}_{\!\:\!t}},
\qquad
\delta \;=\;
\frac{1}{\,\mbox{\small $T$}_{\!\:\!\ell}},
\end{equation}
\mbox{} \vspace{-0.150cm} \\
where $\mbox{\small $T$}_{\!\:\!t}$ denotes the
{\small \sc average transmission time}
and $ \mbox{\small $T$}_{\!\:\!\ell}$ stands for the
mean length of the {\small \sc latent period} 
(the time taken to become infectious, once infected),
which for Covid-19 are typically taken from 10 to 14 days
and from 3 to 5 days, respectively
\cite{Kim2020, Kucharski2020, Lauer2020, %
Rai2021, Singh2021, Vaid2020}.
\!In~(1.1),
$N$ denotes the full size of the susceptible population
initially exposed,
so that we have
$ S(t_0) + E(t_0) + I(t_0) $
$ + R(t_0) + D(t_0) = N \!\:\!$,
where $t_0$ denotes the initial time.
From (1.1), it follows the
{\small \sc conservation law} \\
\mbox{} \vspace{-0.575cm} \\
\begin{equation}
\tag{1.3}
S(t) + E(t) + I(t) + R(t) + D(t) \;\!=\, N,
\qquad
\forall \;\,
t > t_0.
\end{equation}
\mbox{} \vspace{-0.275cm} \\
For the model (1.1) to be useful,
not only the values of $ \beta(t) $ and $r(t)$
must be obtained \linebreak
but also {\small \sc initial values}
for 
$ S(t) $, $\!E(t) $, $\! I(t) $, $\! R(t) \!\;\!$ 
and $ \!\;\!D(t) $
must be provided. 
%
%
Typical available data inform the total 
number $ C(t_i) $ of cases reported 
up until some time moments $ t_i$ and
the total number of deaths, $D(t_i)$.
As $ S(t_i) = N \!\;\!- C(t_i) $,
by (1.3)
we are given the sums
$ \:\!E(t_i) + I(t_i) + R(t_i) $,
but not the individual values
$ E(t_i) $, $ I(t_i) $ and $ R(t_i) $. 
\!This difficulty is addressed in Section~2.
\!A satisfactory solution is important
for obtaining good estimates of reproduction numbers
and herd immunity values.  

With the model (1.1) then completed,
and implementation issues resolved, 
we~are in good position
to estimate herd immunity levels (Section~3). 
Taking ten countries for illustration,
we examine their situation 
in two different periods of the year 2020:
before contention measures were applied
and after the first 75 days of intervention,
when the population was well aware 
of the value of simple protective measures
like hand hygiene and mask wearing.
A few closing remarks are given
in Section~4. 
\newpage
%

%
%
%
%

%
\mbox{} \vspace{-0.550cm} \\
{\bf 2. Implementing the SEIR model} \\
%

Having introduced the {\small SEIR} equations (1.1),
we now describe an implementation of this model
that is suitable for all our needs (and much more). \\
\mbox{} \vspace{-0.700cm} \\

({\em i\/}) {\em assigning a value to
the population parameter $\!\;\!N$} \\
\mbox{} \vspace{-0.750cm} \\

In the case of {\small C}ovid-19,
which was caused by a new virus
({\small SARS}-{\small C}o{\small V}-2),
it is reasonable to assume that the
entire population of the region under consideration
is initially susceptible,
which was done in the code.
In any case,
it turns out that the exact value of
$N$ is not so important
for the short range dynamics
(which accounts for the applications
studied in this paper)
as it proves to be
for long time simulations (Figures 1$a$ and 1$b$).
\nl
%
%
\nl
\mbox{} \vspace{-0.250cm} \\
%
%
%
%
%
%
\mbox{} \hspace{+1.250cm}  
\begin{minipage}[t]{14.600cm}  
\includegraphics[keepaspectratio=true, scale=0.80]{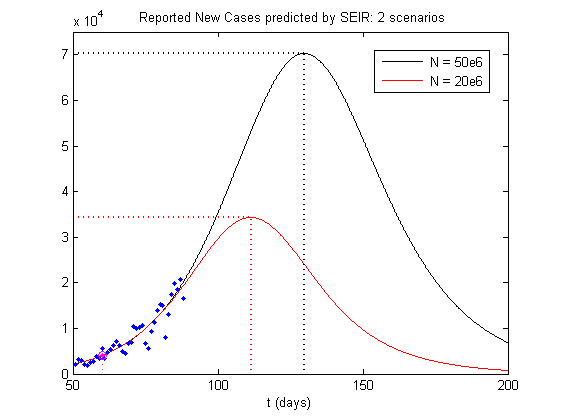}
\end{minipage} \\
\mbox{} \vspace{-0.350cm} \\
\mbox{} \hspace{+0.250cm}
{\footnotesize {\bf Fig.\,1\!\;\!{\em a\/}:}
\begin{minipage}[t]{12.200cm}
{\footnotesize
Prediction by model (1.1)
of the daily number of {\em new cases\/} of Covid-19
expected to be reported in Brazil
between the initial time
$ t = t_0 = 60 $ (April\;25)
and $ t = 200 $ (September\;\!\;\!12),
considering susceptible populations
of $N \!\:\!= 20 $ million (red curve)
and $ N \!\:\!= 50 $ million (black curve).
Note the appreciable difference between \linebreak
the predicted peak values
(34 and 70 thousand, resp.)
\!\:\!and their respective dates, \linebreak
June\:6 and July\:4.
Actual data points are shown in blue.
(\:\!Computed from
data available at the official site\;\!
{\color{blue} \url{https://covid.saude.gov.br}}.) \\
}
\end{minipage}
}
\newpage
\mbox{} \vspace{-0.950cm} \\
%
%
%
%
%
\mbox{} \hspace{+1.250cm}  
\begin{minipage}[t]{14.600cm}  
\includegraphics[keepaspectratio=true, scale=0.80]{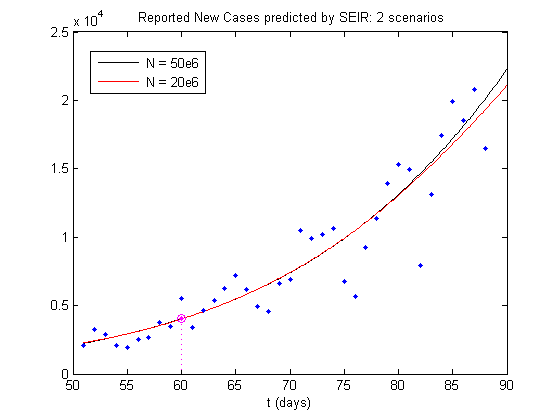}
\end{minipage}
\mbox{} \vspace{-0.350cm} \\
\mbox{} \hspace{+0.350cm}
{\footnotesize {\bf Fig.\,1{\em b\/}:}
\begin{minipage}[t]{12.500cm}
{\footnotesize
Thirty day prediction by model (1.1)
of the daily number of {\em new cases\/} of Covid-19
to be reported in Brazil
between the initial time
$ t = t_0 = 60 $ (04/25)
and $ t = 90 $ (05/25),
considering susceptible exposed populations
of $N \!\;\!= 20 $ million (red curve) \linebreak
and $ N \!\;\!= 50 $ million (black curve).
Note the very close similarity of the two
30D predictions in spite of the appreciable
difference in the values of $N\!\:\!$.
Points shown in blue are the official values
reported (cf.
{\color{blue} \url{https://covid.saude.gov.br}}.)
}
\end{minipage}
\mbox{} \hspace{+1.000cm} 
}
\mbox{} \vspace{-0.500cm} \\

({\em ii\/}) {\em generation of initial data\/}
$S(t_0)$, $ E(t_0) $, $ I(t_0)$, $ R(t_0)$, $ D(t_0)$ \\
\mbox{} \vspace{-0.750cm} \\

Initial values
$S_0, E_0, I_0, R_0, D_0 $
for the five variables are generated
from a starting date $ t_{\!\;\!s}\!\;\!$ on,
which is taken so as to meet some minimum value
chosen
of total reported cases (namely, 100).
\!Denoting by $ C_{r}(t) $
the total amount of reported cases
up to some time $t$,
and letting \mbox{\small EIR}$(t)$
be the sum of the populations
$ E(t) $, $ I(t) $ and $ R(t) $,
we set \\
\mbox{} \vspace{-1.050cm} \\
\begin{equation}
\tag{2.1}
\mbox{\small EIR}(t_s) \,=\,
f_{c} \!\;\!\cdot (\:\!C_{r}(t_s) - D(t_s)),
\end{equation}
\mbox{} \vspace{-0.275cm} \\
where $f_{c} \!\;\!\geq\!\;\! 1 $
denotes a {\small \sc correction factor}
to account for likely underreportings
on the official numbers given.
(\:\!In (2.1), we have neglected possible
underreportings on the number of deaths,
which could of course be similarly accounted for
if desired.)
Again, this factor $f_c$ will not play an important role
in this paper and could be safely ignored,
but it should be carefully considered
in the case of long time predictions.
We have typically taken $ f_c = 5 $.
Having estimated $ \mbox{\small EIR}(t_s) $,
we then set \\
\mbox{} \vspace{+0.000cm} \\
\mbox{} \hspace{+2.250cm}
$ {\displaystyle
E(t_s) \,=\, E_0(t_s)
:=\; a \cdot (1 - b) \!\:\!\cdot \mbox{\small EIR}(t_s)
} $,
\hfill (2.2$a$) \\
\mbox{} \vspace{-0.150cm} \\
\mbox{} \hspace{+2.250cm}
$ {\displaystyle
I(t_s) \,=\, I_{0}(t_s)
:=\; (1 - a) \cdot (1 - b) \!\:\!\cdot \mbox{\small EIR}(t_s)
} $,
\hfill (2.2$b$)
\newpage
\mbox{} \vspace{-0.750cm} \\
\mbox{} \hspace{+2.250cm}
$ {\displaystyle
R(t_s) \,=\, R_{0}(t_s)
:=\; b \cdot \mbox{\small EIR}(t_s)
} $,
\hfill (2.2$c$) \\
\mbox{} \vspace{-0.100cm} \\
\mbox{} \hspace{+2.250cm}
$ {\displaystyle
S(t_s) \,=\,  S_{0}(t_s) :=\,
N -\, \bigl(\;\! E(t_s) + I(t_s) + R(t_s) + D(t_s) \:\!\bigr)
} $,
\hfill (2.2$d$) \\
\mbox{} \vspace{-0.050cm} \\
where $ \!\;\!\;\!a =\!\;\!\;\! \mbox{\small $T$}_{\!\:\!\ell} /
(\mbox{\small $T$}_{\!\:\!\ell} +\!\;\!\;\!\mbox{\small $T$}_{\!\:\!t}) $
and $\!\;\!\;\! b \!\;\!\;\!=\!\;\!\;\! 0.30 $.
%
%
Although (2.2) above might seem reasonable, \linebreak
the expressions (2.2$a$)\;\!-\;\!(2.2$c$)
are nevertheless arbitrary and will be probably in error.  
However, 
all the errors will eventually fade away (Figure~2)
as we compute more values
$ S_0(t_0), E_0(t_0), I_0(t_0), R_0(t_0), D_0(t_0) $
at later initial times
\mbox{$ \:\!t_0  \!\:\! = t_s \!\:\!+\!\:\! 1,\!\:\!..., t_{\mbox{}_{\!\:\!F}} \!\;\!$},
where \mbox{$ \:\! t_{\mbox{}_{\!\:\!F}} \!\:\!$}
stands for the final (i.e., most recent)
day of reported data available. 
This can be done as follows. 
For each $ t_0 $,
the solution of the equations (1.1)
with the previously obtained initial data
at \mbox{\:\!$ t_0 \!\:\!-\!\:\! 1 $}
is computed on the interval
\mbox{$ J(t_0) = [\,t_0\!\:\! -\!\:\! 1,\;\! t_1\;\!] $},
\mbox{$ t_1 \!\:\!=\;\! \min\;\!
\{\;\! t_0\!\:\! -\!\:\! 1\!\;\! + d_0, \, t_{\mbox{}_{\!\:\!F}} \!\;\!\} $},
with constant parameters
\mbox{$ \beta = \beta_0(t_0 \!\:\!-\!\:\! 1)$},
\mbox{$ r = r_0(t_0 \!\:\!-\!\:\! 1)$}
determined so that
the computed values for
$C_r(t) $, $ D(t) $
best fit the reported data
for these variables
on \mbox{$ [\;\!t_0, \;\!t_{1}\:\! ] $}
in the sense of {\small \sc least\;squares}
\cite{Martcheva2015}.
\!(\:\!Here,
\mbox{$ d_0 \!\;\!\in [\,2, \;\!10\;\!] $}
is chosen according to the data regularity.)
Once this solution $(S,E,I,R,D)(t)$
is obtained,
we set
$ S_0(t_0) \!:= S(t_0) $,
$ E_0(t_0) \!:= E(t_0) $,
$ I_0(t_0) \!:= I(t_0) $,
$ R_0(t_0) \!:= R(t_0) $,
$ D_0(t_0) \!:= D(t_0) $
and move on to the
next time level
\mbox{$\:\!t_0 \!\:\!+\!\:\! 1$},
repeating the procedure until
\mbox{$t_{\mbox{}_{\!\:\!F}}$}\!\:\!
is reached. \\
\nl
%
%
%
%
%
%
\mbox{} \hspace{+1.250cm}  
\begin{minipage}[t]{14.600cm}  
\includegraphics[keepaspectratio=true, scale=0.80]
{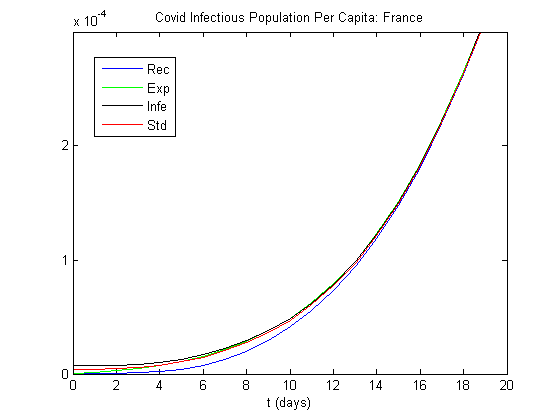}
\end{minipage}
\mbox{} \vspace{-0.550cm} \\
\mbox{} \hspace{-0.150cm}
{\footnotesize {\bf Fig.\,2{\em a\/}:}
\begin{minipage}[t]{13.700cm}
{\footnotesize
Self\:\!-\:\!correction in the initialization procedure 
({\em ii\/}) to generate 
$ S_0(t_0) $, $ E_0(t_0) $, $ I_0(t_0) $
and $ R_0(t_0) $ 
for $ t_0 = t_{s}, t_{s}+1 $,\;\!..., $ \!\;\!t_{F}$,
showing above the case of $ I_0(t_0)/N $ in France.
At time $ t_{s} = 0 $,  
the day of 100 total cases reported (29/02/2020),
four very different sets of values 
$ \{\:\! E_0(t_s), \:\!I_0(t_s), \:\!R_0(t_s)\:\!\} $ 
are considered: our standard choice (2.2$a$)\:\!-\:\!(2.2$c$),
shown in red; $E_0(t_s) = I_0(t_s) = 0.05 \times \mbox{EIR}(t_s) $, 
$ R_0(t_s) = 0.90 \times \mbox{EIR}(t_s) $, shown in blue;  
$ E_0(t_s) = \mbox{EIR}(t_s) $,
$ I_0(t_s) = R_0(t_s) = 0 $, shown in green;
and 
$ I_0(t_s) = \mbox{EIR}(t_s) $,
$ E_0(t_s) = R_0(t_s) = 0 $, in black.
For $t_0 > t_s + 15 $, 
all four initializations produce essentially the {\em same\/} values
for $ I_0(t_0) $. 
}
\end{minipage}
\mbox{} \hspace{+1.000cm} 
}
\mbox{} \vspace{-1.400cm} \\
%
%
%
%
%
%
\mbox{} \hspace{+1.250cm}  
\begin{minipage}[t]{14.600cm}  
\includegraphics[keepaspectratio=true, scale=0.80]
{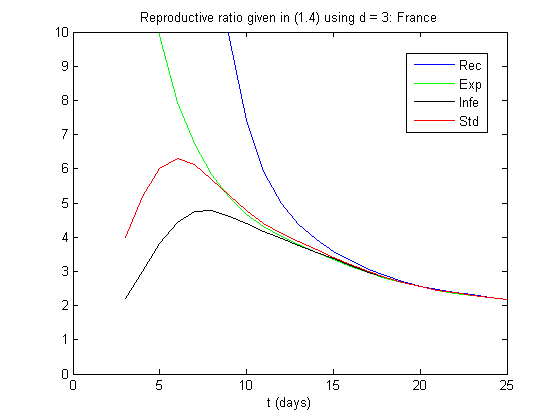}
\end{minipage}
\mbox{} \vspace{-0.600cm} \\
\mbox{} \hspace{-0.150cm}
{\footnotesize {\bf Fig.\,2{\em b\/}:}
\begin{minipage}[t]{13.700cm}
{\footnotesize
Self\:\!-\:\!correction in the values of
$ R_t = I_0(t+3)/I_0(t-3) $
induced by the correction
of $ E_0(t_0) $, $ I_0(t_0) $ and $ R_0(t_0) $ 
exhibited in Fig.\,2$a$,
considering the same four
initializations at $ t_s \!\;\!= 0 $. 
%
}
\end{minipage}
\mbox{} \hspace{+1.000cm} 
}
\mbox{} \vspace{-0.525cm} \\

({\em iii\/}) {\em computing the solution on
some final interval} $ [\,t_0, \;\!\mbox{\small $T$}\;\!] $
\!({\small \sc prediction phase}) \\
\mbox{} \vspace{-0.650cm} \\

Having completed the previous steps,
we can address the possibility of
{\em prediction}.
Although this is not important for our present goals,
it is included for completeness.
Choosing an initial time
\mbox{$\:\!t_0 \!\;\!\in (\:\! t_{s}, \;\!t_{\mbox{}_{\!\:\!F}}] $},
we then take the initial values \\
\mbox{} \vspace{-0.650cm} \\
\begin{equation}
\notag
S(t_0) = S_{0}(t_0),
\;
E(t_0) = E_{0}(t_0),
\;
I(t_0) = I_{0}(t_0),
\;
R(t_0) = R_{0}(t_0),
\;
D(t_0) = D_{0}(t_0).
\end{equation}
\mbox{} \vspace{-0.250cm} \\
In order to predict the values of
the variables
$ S(t), E(t), I(t), R(t), D(t) $
for $ t > t_0 $,
it is important to have
good estimates for the
evolution of the key parameters
$ \beta(t) $ and $ r(t) $
beyond $t_0$.
This may be a particularly computationally intensive
part of the algorithm.
%
\!Such estimates can be
given in the form \\
\mbox{} \vspace{-0.600cm} \\
\begin{equation}
\tag{2.3$a$}
\beta(t) \,=\; \beta_0 \;\!+\, a_{\beta} \,
e^{\mbox{\footnotesize $ -\,\lambda_{\beta}(\:\!t - \:\!t_0)$}}
\end{equation}
\mbox{} \vspace{-1.050cm} \\
\begin{equation}
\tag{2.3$b$}
r(t) \,=\; r_0 \;\!+\, a_{r} \,
e^{\mbox{\footnotesize $ -\,\lambda_{r}(\:\!t - \:\!t_0)$}}
\end{equation}
\mbox{} \vspace{-0.375cm} \\
where
$\beta_0, a_{\beta}, \lambda_{\beta}, r_0, a_{r}, \lambda_{r}
\!\;\!\in \mathbb{R} $
are determined
so as to minimize the maximum size
of weighted {\small \sc relative errors\/}
in the computed values for
$C_{r}(t), D(t) $
as compared to the official data reported
for these variables
on some previous interval
\mbox{$ [\;\! t_{0}\!\:\!-\!\:\!\tau_{0}, \;\!t_{0} \:\!] $}
(weighted {\small \sc Chebycheff} {\small \sc problem})
for some chosen
$ \tau_0 > 0 $
(usually, $ 20 \leq \tau_0 \leq 30 $).
This problem is solved iteratively
starting with an initial guess
obtained from the analysis of the
previous values $ \beta_0(t), r_0(t) $
computed in the step ({\em ii\/}) above.
\!\mbox{The result} is illustrated
in Figure~3 for the case of $ \beta(t) $,
with similar considerations for $ r(t)$.

\newpage
\mbox{} \vspace{-1.500cm} \\
%
%
%
%
%
\mbox{} \vspace{-0.400cm} \\
\mbox{} \hspace{+1.500cm}
\begin{minipage}[t]{14.600cm}  
\includegraphics[keepaspectratio=true, scale=0.75]{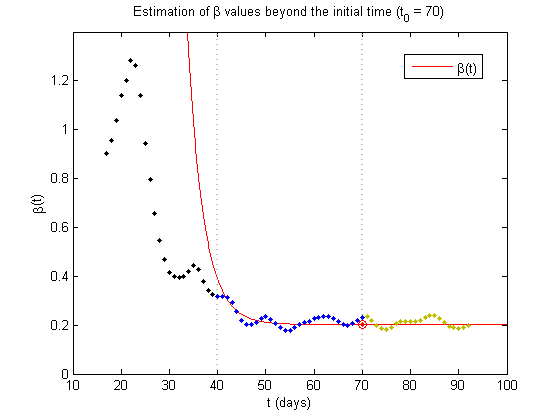} 
\end{minipage}
\mbox{} \vspace{-0.650cm} \\
\mbox{} \hspace{-0.250cm}
{\footnotesize {\bf Fig.\,3:}
\begin{minipage}[t]{14cm} 
{\footnotesize
Estimation of future values of
the transmission parameter $ \beta(t) $
beyond the initial time $ t_{0} $
$ = 70 $ (05/05/2020)
for the outbreak of Covid-19 in Brazil,
assuming the basic form (2.3$a$),
after solving the
Chebycheff problem (red curve).
The data points in the interval
\mbox{$ [\;\!40, 70\;\!] $},
shown here in blue,
are values of the function $ \beta_0(t) $
computed in step ({\em ii\/}),
which are used to obtain the
first approximation to $ \beta(t) $.
Values of $ \beta_0(t)$
previous to $ t = 40 $ (04/05/2020),
shown in black, are disregarded.
\!The golden points beyond $ t_0 \!\;\!= 70 $
are future values of $ \beta_0(t) $, \linebreak
not available on 05/05/2020,
displayed to allow comparison
with the predicted values $ \beta(t) $. \linebreak
}
\end{minipage}
}
\nl
\mbox{} \vspace{-0.500cm} \\
Once $ \beta(t) $, $ r(t) $
have been obtained,
the equations (1.1) are finally solved
(Figure~4). \linebreak
\mbox{} \vspace{-0.550cm} \\
%

%
%
%
%
\mbox{} \vspace{-0.600cm} \\
\mbox{} \hspace{+1.500cm}
\begin{minipage}[t]{14.600cm}  
\includegraphics[keepaspectratio=true, scale=0.80]{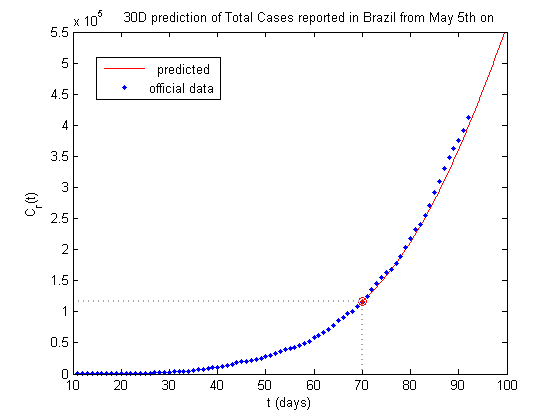}
\end{minipage}
\mbox{} \vspace{-0.600cm} \\
\mbox{} \hspace{-0.250cm}
\nopagebreak
{\footnotesize {\bf Fig.\,4:}
\begin{minipage}[t]{14cm}
{\footnotesize
Computation of
$ C_{r}(t) = \bigl( E(t) + I(t) + R(t) \bigr)/f_{c}
+ D(t) $ for $\;\! t > t_0 = 70 $ (05/05/2020),
with initial data
$ C_{r}(t_0) = \bigl(E_0(t_0) + I_0(t_0) + R_0(t_0) \bigr)/f_{c} + D_0(t_0) $,
after obtaining $ \beta(t) $, $ r(t) $ -- see Fig.\,3 for $ \beta(t)$.
The numerical solution of equations (1.1)
is easily obtained by any method.
}
\end{minipage}
}
\mbox{} \vspace{-1.250cm} \\
%
%
%
%

{\bf 3. Herd Immunity Estimates for Covid-19} \\
\mbox{} \vspace{-0.575cm} \\

In this section
we use the {\small SEIR} 
algorithm developed 
in Section~2
to estimate 
immunization levels
for a given population
that could protect it
against a {\small C}ovid-19 outbreak.
Since the results
depend strongly 
on the
transmission rates
that would likely be observed,
two distinct scenarios
are considered: 
the case of an unsuspecting population
caught off guard against the disease,
and a warned population that 
takes some protective measures. 
Whether or not an outbreak
will happen can be answered
by examining  
the values of
reproduction numbers
associated with the infected population. 
Summing up the second and third equations in (1.1),
we have \\
\mbox{} \vspace{-0.600cm} \\
\begin{equation}
\tag{3.1}
\frac{\;\!d\;\!\mathbb{I}\;\!}{d\;\!t}
\;\!\;\!=\; \alpha(t) \;\!I(t),
\quad \;\;\,
\alpha(t) \,=\;\!\;\!
\beta(t) \, \frac{\:\!S(t)\;\!}{N} \,-\;\!\;\! r(t) -\:\! \gamma,
\end{equation}
\mbox{} \vspace{-0.250cm} \\
where $ \mathbb{I}(t) = E(t) + I(t) $ is
the infected population at time $t$, 
$ \gamma $ is given in (1.2)
and $ \beta(t), r(t) $ 
are the transmission and lethality rates
of the disease,
which are obtained 
as described in Section~2.
Whether or not 
$ \mathbb{I}(t) $ decreases
depends on whether or not
$ \alpha(t) < 0 $,
that is, on
whether the corresponding
reproduction number \\
\mbox{} \vspace{-0.550cm} \\
\begin{equation}
\tag{3.2}
{\cal R}_{t}
\,=\; 
\frac{\,\beta \!\;\!\cdot \!\;\!S \:\!/ N\:\!}
{r + \gamma}
\end{equation}
\mbox{} \vspace{-0.225cm} \\
is smaller than the threshold value 1 or not. 
\!Assuming that a fraction $p$ 
of the population has been immunized, 
so that $ S/N = 1 - p $, 
the condition
$ {\cal R}_t \!\;\!< 1 $ 
becomes \\
\mbox{} \vspace{-0.550cm} \\
\begin{equation}
\tag{3.3}
p \;>\: \hat{p} \:\!:=\;
1 \,-\; 
\frac{1}{\;\mathbb{R}_{t}},
\qquad
\mathbb{R}_{t} \,=\; 
\frac{\beta}{\;\!r + \gamma \;\!}
\end{equation}
\mbox{} \vspace{-0.200cm} \\
(see e.g.\;\!\;\!\cite{Martcheva2015}, p.\;\!\;\!217),
and the problem then reduces to obtaining 
reliable estimates of the basic reproductive ratios
$ \:\!\mathbb{R}_{t} \!\;\!$
given in~(3.3). \\
\mbox{} \vspace{-0.450cm} \\
%
%
%
%

({\em i\/}) 
{\em the case of an unsuspecting population 
caught \!\;\!off guard\/}:  
$ 80\:\!\mbox{\small \%} < \hat{p} < 90\:\!\mbox{\small \%} $ \\ 
\mbox{} \vspace{-0.575cm} \\

That is,
the population
is not aware of the presence of any infected individuals,
and contention or hygienic measures
are not being observed.
\!This was the case
of most countries around the world
in the first weeks of their 2020 covid-19 outbreak. 
We can then obtain
the $ \mathbb{R}_t \!\;\! $ values needed
to estimate $ \hat{p} \!\;\!\!\;\!\;\!$
by looking at what happened  
there 
in the days 
before the application of intervention measures,
neglecting the initial results
as they are
subject to initialization errors
(Section~2).
This is illustrated in 
Figure~5 below,
where we assumed
$ \gamma = 0.1 $,
i.e., 
a transmission period of 10 days,
and
$ \delta = 1/3 $, that is,
a latency period of 3 days, 
cf.\;(1.2).  

%
%
%

%
%
%
%
%
%
%
%
\mbox{} \hspace{-3.125cm} 
\begin{minipage}[t]{9.10cm}  
\includegraphics[keepaspectratio=true, scale=0.52]{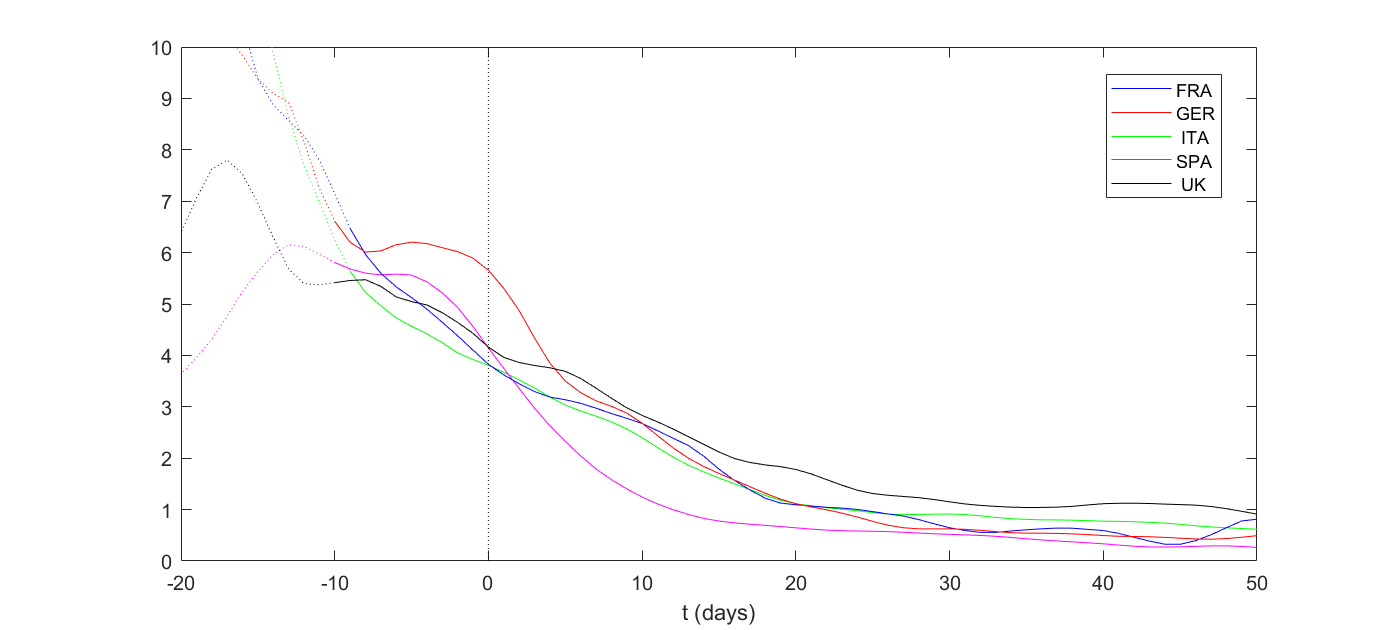}
\end{minipage}
\mbox{} \vspace{-0.450cm} \\
\nl
\mbox{} \hspace{+0.175cm} 
{\footnotesize {\bf Fig.\,5{\em a\/}:}
\begin{minipage}[t]{12.950cm}
{\footnotesize
Time history of the reproductive ratios \!\;\!\;\!$\mathbb{R}_{t} \!\!\;\!\;\!$
defined by (3.3)
in five European countries 
(France, Germany, Italy, Spain and UK)
about the time of their first
intervention measures 
against Covid-19 ($t = 0$),
assuming $T_t = 10 $ and $ T_{\ell} = 3 $. 
The initial dashed parts of the curves
correspond to  
the initialization phase of the algorithm
before self-correction
(Section~2),
with the values shown in these parts
being disregarded.  
}
\end{minipage}
}
\nl
%

%
%
%
%
\mbox{} \hspace{-3.125cm} 
\begin{minipage}[t]{9.10cm}  
\includegraphics[keepaspectratio=true, scale=0.52]{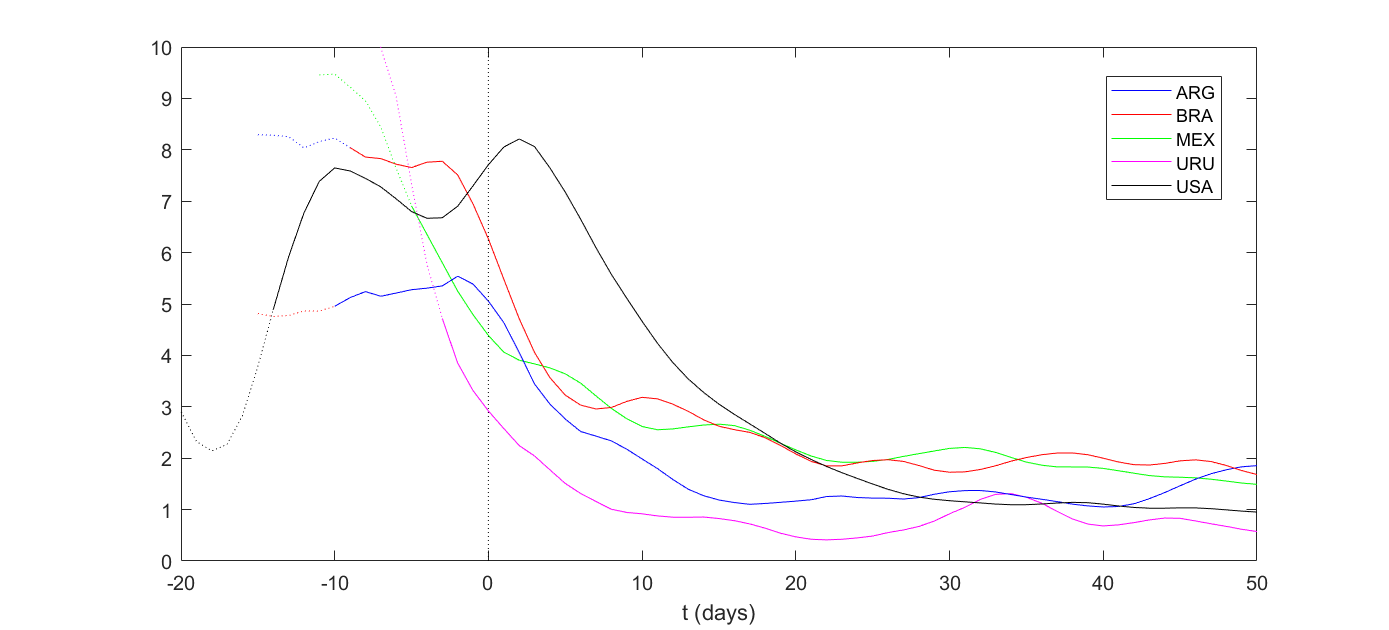}
\end{minipage}
\mbox{} \vspace{-0.450cm} \\
\nl
\mbox{} \hspace{+0.175cm} 
{\footnotesize {\bf Fig.\,5{\em b\/}:}
\begin{minipage}[t]{12.950cm}
{\footnotesize
Time history of the reproductive ratios \!\;\!\;\!$\mathbb{R}_{t} \!\!\;\!\;\!$
defined by (3.3)
in five American countries 
(Argentina, Brazil, Mexico, Uruguay and US)
about the time of their first
intervention measures 
against Covid-19 ($t = 0$),
assuming again $T_t = 10 $ and $ T_{\ell} = 3 $.  
As in Fig.\,5$a$, the dashed parts
correspond to  
the initialization phase of the algorithm
before self-correction
(Section~2),
with the respective values
disregarded. 
All computations based on
covid data available 
for the 10 countries at 
\href{https://www.worldometers.info/coronavirus/}%
{{\color{blue} worldometers/coronavirus}}.
}
\end{minipage}
}
\mbox{} \vspace{-0.550cm} \\
Observing the numerical ranges 
of the ratios
$ \mathbb{R}_{t} \!\;\!$
in the region $ t \leq 0 $,
pictured
in Figure~5,
we obtain the results displayed
in the second and third columns 
of {\small T}able~{\!\:\!1}. 
Larger transmission or latency periods
yield larger 
$ \mathbb{R}_{t} \!\:\!$
values
in this region,
leading to
larger estimates 
of herd immuntity levels.
\!For example,
taking $T_t \!\;\!= 14 $
and $ T_{\ell} = 5 $
we obtain,
repeating the procedure,
the results given
in the last column
of Table~{\!\:\!1}. 
\nl
\mbox{} \vspace{-0.150cm} \\
\mbox{} \hspace{+0.200cm} 
{\small
\mbox{} \hspace{-0.450cm} 
\begin{tabular}{||c|c|c|c||}
\hline \hline
{\bf Country} & \,\mbox{\boldmath $ \mathbb{R}_{t} $} {\bf range}
\mbox{\footnotesize \bf \;\!(Case I\:\!)}\, 
& \,\mbox{\boldmath $ \hat{p} $}\:\! {\bf range} 
\mbox{\footnotesize \bf \;\!(Case I\:\!)}\,
& \,\mbox{\boldmath $ \hat{p} $}\:\! {\bf range} 
\mbox{\footnotesize \bf \;\!(Case II\:\!)}\, \\
\hline
Argentina & $4.95 < \:\!\mathbb{R}_{t} < 5.55 \;$ 
& $\; 79\:\!\mbox{\small \%} \:\!<\;\! \hat{p} 
  \:\!<\:\! 82\:\!\mbox{\small \%} \;$ 
& $\; 85\:\!\mbox{\small \%} \:\!<\;\! \hat{p} 
  \:\!<\:\! 89\:\!\mbox{\small \%} \;$ \\
\hline
Brazil & $\; 6.26 < \:\!\mathbb{R}_{t} < 8.05 \;$
& $\; 84\:\!\mbox{\small \%} \:\!<\;\! \hat{p} 
  \:\!<\:\! 88\:\!\mbox{\small \%} \;$ 
& $\; 90\:\!\mbox{\small \%} \:\!<\;\! \hat{p} 
  \:\!<\:\! 93\:\!\mbox{\small \%} \;$ \\
\hline
France & $\; 3.82 < \:\!\mathbb{R}_{t} < 6.47 \;$
& $\; 73\:\!\mbox{\small \%} \:\!<\;\! \hat{p} 
  \:\!<\:\! 85\:\!\mbox{\small \%} \;$ 
& $\; 83\:\!\mbox{\small \%} \:\!<\;\! \hat{p} 
  \:\!<\:\! 91\:\!\mbox{\small \%} \;$ \\
\hline
Germany & $\; 5.65 < \:\!\mathbb{R}_{t} < 6.62 \;$
& $\; 82\:\!\mbox{\small \%} \:\!<\;\! \hat{p} 
  \:\!<\:\! 85\:\!\mbox{\small \%} \;$ 
& $\; 89\:\!\mbox{\small \%} \:\!<\;\! \hat{p} 
  \:\!<\:\! 92\:\!\mbox{\small \%} \;$ \\
\hline
Italy & $ 3.80 < \:\!\mathbb{R}_{t} < 5.65 $ 
& $ 73\:\!\mbox{\small \%} \:\!<\;\! \hat{p} 
  \:\!<\:\! 83\:\!\mbox{\small \%} $ 
& $\; 83\:\!\mbox{\small \%} \:\!<\;\! \hat{p} 
  \:\!<\:\! 90\:\!\mbox{\small \%} \;$ \\
\hline
Mexico & $ 4.38 < \:\!\mathbb{R}_{t} < 6.91 $ 
& $ 77\:\!\mbox{\small \%} \:\!<\;\! \hat{p} 
  \:\!<\:\! 86\:\!\mbox{\small \%} $ 
& $\; 86\:\!\mbox{\small \%} \:\!<\;\! \hat{p} 
  \:\!<\:\! 92\:\!\mbox{\small \%} \;$ \\
\hline
Spain & $ 4.14 < \:\!\mathbb{R}_{t} < 5.81 $ 
& $ 76\:\!\mbox{\small \%} \:\!<\;\! \hat{p} 
  \:\!<\:\! 83\:\!\mbox{\small \%} $ 
& $\; 85\:\!\mbox{\small \%} \:\!<\;\! \hat{p} 
  \:\!<\:\! 90\:\!\mbox{\small \%} \;$ \\
\hline
United Kingdom & $ 4.16 < \:\!\mathbb{R}_{t} < 5.48 $ 
& $ 75\:\!\mbox{\small \%} \:\!<\;\! \hat{p} 
  \:\!<\:\! 82\:\!\mbox{\small \%} $ 
& $\; 84\:\!\mbox{\small \%} \:\!<\;\! \hat{p} 
  \:\!<\:\! 89\:\!\mbox{\small \%} \;$ \\
\hline
Uruguay & $ 2.91 < \:\!\mathbb{R}_{t} < 4.72 $ 
& $ 65\:\!\mbox{\small \%} \:\!<\;\! \hat{p} 
  \:\!<\:\! 79\:\!\mbox{\small \%} $ 
& $\; 79\:\!\mbox{\small \%} \:\!<\;\! \hat{p} 
  \:\!<\:\! 88\:\!\mbox{\small \%} \;$ \\
\hline
United States & $ 4.88 < \:\!\mathbb{R}_{t} < 7.71 $ 
& $ 79\:\!\mbox{\small \%} \:\!<\;\! \hat{p} 
  \:\!<\:\! 88\:\!\mbox{\small \%} $ 
& $\; 86\:\!\mbox{\small \%} \:\!<\;\! \hat{p} 
  \:\!<\:\! 93\:\!\mbox{\small \%} \;$ \\
\hline \hline
\end{tabular}
\nl
}
\nl
\mbox{} \vspace{-0.200cm} \\
\mbox{} \hspace{+0.200cm} 
{\small {\bf Table\;1:}\!
\begin{minipage}[t]{12.50cm}
{\small
Estimates of herd immunity levels 
$ \;\!\hat{p} = 1 - 1/\mathbb{R}_t $
for {\footnotesize C}ovid-19 
in the early days of the epidemic,
before the application of contention measures.
Case I corresponds to
average latency and transmission 
periods of 3 and 10 days,
respectively,
while Case II assumes
longer periods of
5 and 14 days. 
Computation of 
transmission and lethality rates 
for each country
uses the {\footnotesize SEIR} algorithm
above
and covid data
available at
\href{https://www.worldometers.info/coronavirus/}%
{{\color{blue} worldometers/coronavirus}}.
}
\end{minipage}
}
%
%
%
%
%
%
%
\nl
\nl

The values obtained
are suggestive of a basic range
$ 80\:\!\mbox{\small \%} < \hat{p} < 90\;\!\mbox{\small \%} $
in case ({\em i\/}). \\
\mbox{} \vspace{-0.350cm} \\
%
%
%
%

%
({\em ii\/}) 
{\em the case of a wary population 
observing basic measures\/}:  
$ 50\:\!\mbox{\small \%} < \hat{p} < 60\:\!\mbox{\small \%} $ \\ 
\mbox{} \vspace{-0.600cm} \\

By basic measures we mean
simple hygienic procedures
like washing hands~and
wearing masks, and perhaps
some occasional social distancing,
adopted by the majority of 
the population.     
This behavior was observed
in most countries after
the removal or relaxing of stricter
intervention rules
like severe mobility restrictions,
curfew or lockdown measures. 
Estimates for $ \mathbb{R}_{t}\!\:\! $
in such case can then be obtained
by looking at what happened in the year 2020
to this indicator in various countries
after the initial contention measures 
have been relaxed or removed
and searching for
the maximum values 
in this period (Figure~6). 
The results are shown
in Table~2
and seem indicative of 
a basic range
$\:\! 50\:\!\mbox{\small \%} < \hat{p} < 60\:\!\mbox{\small \%} $
\:\!in this scenario.
%
%
%
%
%
%
%
%
%
%

%
\mbox{} \vspace{-0.950cm} \\
%

%
%
%
%
%
%
%
%
\mbox{} \hspace{-3.000cm} 
\begin{minipage}[t]{9.10cm}  
\includegraphics[keepaspectratio=true, scale=0.52]{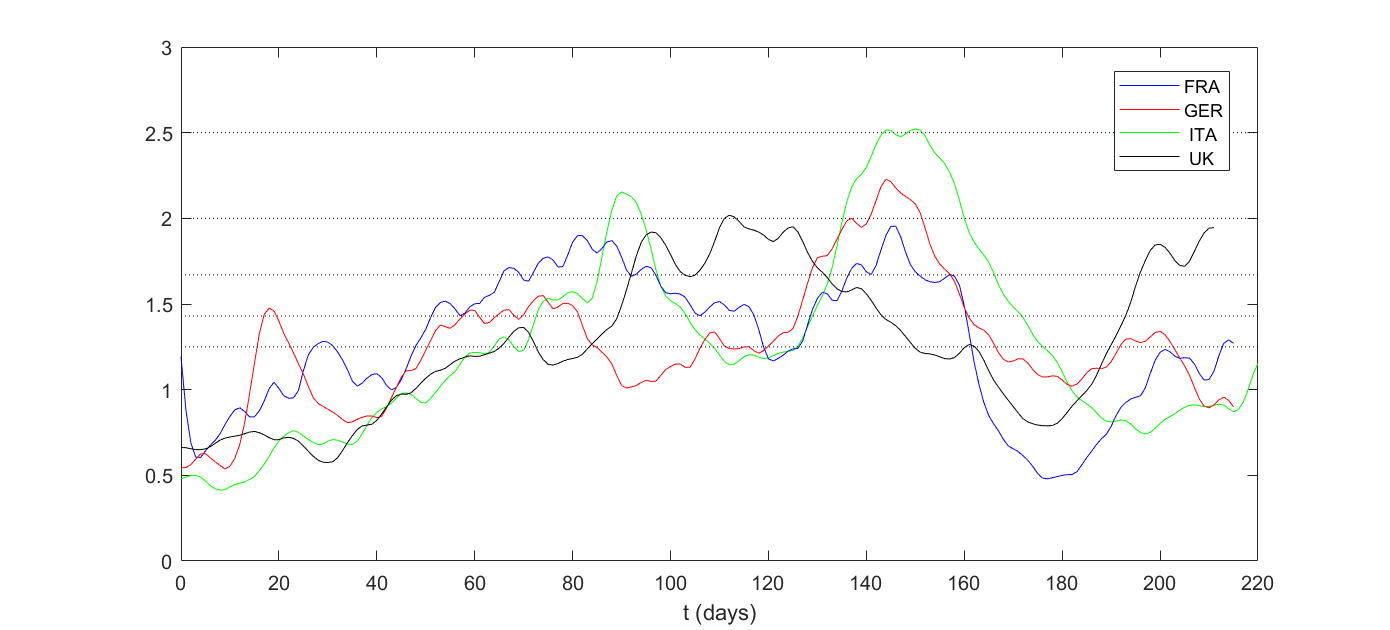}
\end{minipage}
\mbox{} \vspace{-0.450cm} \\
\nl
\mbox{} \hspace{+0.175cm} 
{\footnotesize {\bf Fig.\,6{\em a\/}:}
\begin{minipage}[t]{13.00cm}
{\footnotesize
Time history in 2020 
of the reproductive 
numbers (3.3)
in four European countries 
(France, Germany, Italy and UK)
after 75 days ($t = 0$)
following the first intervention measures,
assuming $ T_{\ell} = 3 $ 
and $ T_t = 10 $, see (1.2). 
The levels $ 1.25 $,\;$ 1.43 $,\;$ 1.67 $,
$ 2.00 $ and $ 2.50 $ 
(dashed lines)
correspond to 
\mbox{$ \;\!\hat{p} \;\!=\;\! 20\:\!\mbox{\footnotesize \%}$, 
$\!\;\! 30\:\!\mbox{\footnotesize \%} $, 
$\!\;\! 40\:\!\mbox{\footnotesize \%} $, 
$\!\;\! 50\:\!\% \:\!$ and 
$ 60\:\!\% $}.
\!All computations based on
covid data available 
for these countries at 
\href{https://www.worldometers.info/coronavirus/}%
{{\color{blue} worldometers/coronavirus}}.
}
\end{minipage}
}
\mbox{} \vspace{-0.250cm} \\
%

%
%
%
%
\mbox{} \hspace{-2.975cm} 
\begin{minipage}[t]{9.10cm}  
\includegraphics[keepaspectratio=true, scale=0.52]{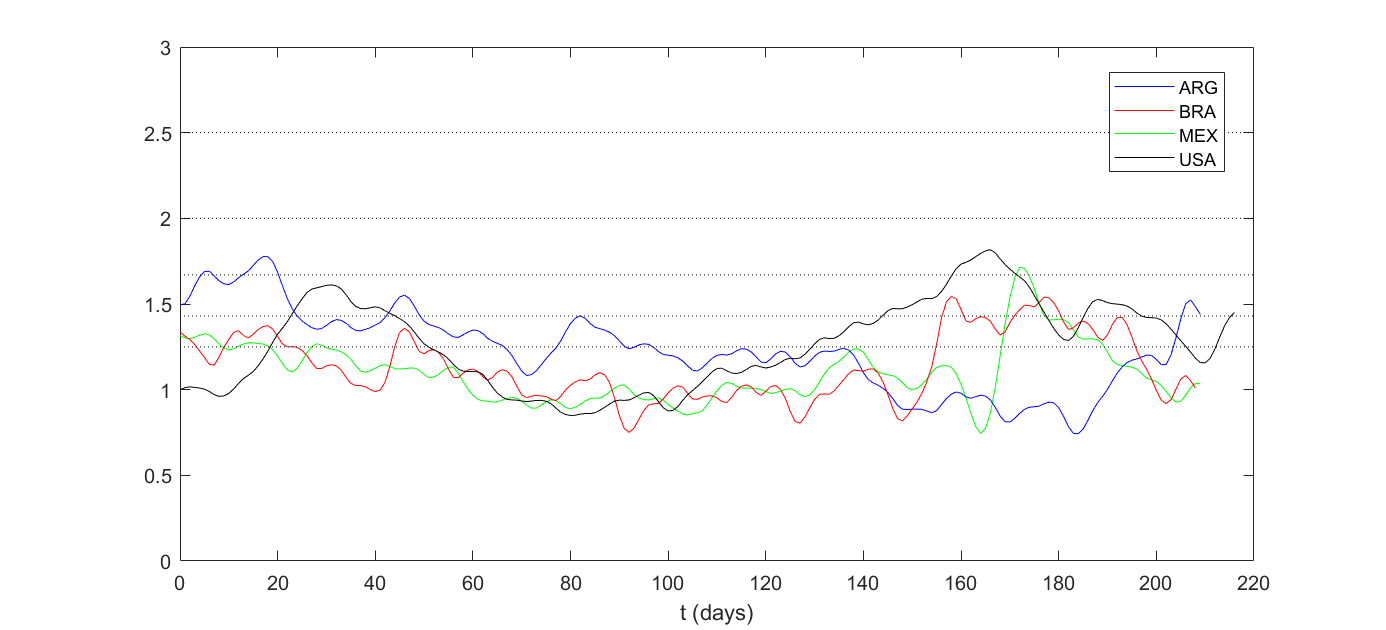}
\end{minipage}
\nl
\mbox{} \hspace{+0.175cm} 
{\footnotesize {\bf Fig.\,6{\em b\/}:}
\begin{minipage}[t]{13.00cm}
{\footnotesize
Time history in 2020 
of the reproductive 
numbers (3.3)
in four American countries 
(Argentina, Brazil, Mexico and US)
after 75 days ($t = 0$)
following the first intervention measures,
assuming again $ T_{\ell} = 3 $ 
and $ T_t = 10 $. 
The levels $ 1.25 $,\;$ 1.43 $,\;$ 1.67 $,
$ 2.00 $ and $ 2.50 $ 
(dashed lines)
correspond to 
\mbox{$ \;\!\hat{p} \;\!=\;\! 20\:\!\mbox{\footnotesize \%}$, 
$\!\;\! 30\:\!\mbox{\footnotesize \%} $, 
$\!\;\! 40\:\!\mbox{\footnotesize \%} $, 
$\!\;\! 50\:\!\% \:\!$ and 
$ 60\:\!\% $}.
\!All computations based on
covid data available 
for these countries at 
\href{https://www.worldometers.info/coronavirus/}%
{{\color{blue} worldometers/coronavirus}}.
}
\end{minipage}
}
%
%
%

%
\nl
\mbox{} \vspace{-0.150cm} \\
\mbox{} \hspace{+0.200cm} 
{\small
\mbox{} \hspace{+0.050cm} 
\begin{tabular}{||c|c|c|c|c||}
\hline \hline
{\bf Country} & \,\mbox{\boldmath $ \mathbb{R}_{t} $} 
\mbox{\footnotesize \bf (Case I\:\!)}\;\! 
& \,\mbox{\boldmath $ \hat{p} $}\:\!
\mbox{\footnotesize \bf (Case I\:\!)}\,
& \,\mbox{\boldmath $ \mathbb{R}_{t} $} 
\mbox{\footnotesize \bf (Case II\:\!)}\;\! 
& \,\mbox{\boldmath $ \hat{p} $}\:\!  
\mbox{\footnotesize \bf (Case II\:\!)}\;\! \\
\hline
Argentina 
& $\; \mathbb{R}_{t} < 1.78 \;$ 
& $\; \hat{p} \:\!<\:\! 44 \:\!\mbox{\small \%} \;$
& $\; \mathbb{R}_{t} < 2.19 \;$ 
& $\; \hat{p} \:\!<\:\! 55 \:\!\mbox{\small \%} \;$ \\
\hline
Brazil 
& $\; \mathbb{R}_{t} < 1.55 \;$ 
& $\; \hat{p} \:\!<\:\! 36 \:\!\mbox{\small \%} \;$
& $\; \mathbb{R}_{t} < 1.83 \;$ 
& $\; \hat{p} \:\!<\:\! 46 \:\!\mbox{\small \%} \;$ \\
\hline
France
& $\; \mathbb{R}_{t} < 1.96 \;$ 
& $\; \hat{p} \:\!<\:\! 49 \:\!\mbox{\small \%} \;$
& $\; \mathbb{R}_{t} < 2.39 \;$ 
& $\; \hat{p} \:\!<\:\! 58 \:\!\mbox{\small \%} \;$ \\
\hline
Germany 
& $\; \mathbb{R}_{t} < 2.23 \;$ 
& $\; \hat{p} \:\!<\:\! 56 \:\!\mbox{\small \%} \;$
& $\; \mathbb{R}_{t} < 2.83 \;$ 
& $\; \hat{p} \:\!<\:\! 65 \:\!\mbox{\small \%} \;$ \\
\hline
Italy
& $\; \mathbb{R}_{t} < 2.53 \;$ 
& $\; \hat{p} \:\!<\:\! 61 \:\!\mbox{\small \%} \;$
& $\; \mathbb{R}_{t} < 3.42 \;$ 
& $\; \hat{p} \:\!<\:\! 71 \:\!\mbox{\small \%} \;$ \\
\hline
Mexico 
& $\; \mathbb{R}_{t} < 1.72 \;$ 
& $\; \hat{p} \:\!<\:\! 42 \:\!\mbox{\small \%} \;$
& $\; \mathbb{R}_{t} < 1.79 \;$ 
& $\; \hat{p} \:\!<\:\! 45 \:\!\mbox{\small \%} \;$ \\
\hline
Spain
& $\; \mathbb{R}_{t} < 2.40 \;$ 
& $\; \hat{p} \:\!<\:\! 59\:\!\mbox{\small \%} \;$
& $\; \mathbb{R}_{t} < 2.99 \;$ 
& $\; \hat{p} \:\!<\:\! 67\:\!\mbox{\small \%} \;$ \\
\hline
United Kingdom
& $\; \mathbb{R}_{t} < 2.02 \;$ 
& $\; \hat{p} \:\!<\:\! 51 \:\!\mbox{\small \%} \;$
& $\; \mathbb{R}_{t} < 2.52 \;$ 
& $\; \hat{p} \:\!<\:\! 61\:\!\mbox{\small \%} \;$ \\
\hline
Uruguay
& $\; \mathbb{R}_{t} < 2.75 \;$ 
& $\; \hat{p} \:\!<\:\! 64 \:\!\mbox{\small \%} \;$
& $\; \mathbb{R}_{t} < 3.04 \;$ 
& $\; \hat{p} \:\!<\:\! 68 \:\!\mbox{\small \%} \;$ \\
\hline
United States
& $\; \mathbb{R}_{t} < 1.82 \;$ 
& $\; \hat{p} \:\!<\:\! 46 \:\!\mbox{\small \%} \;$
& $\; \mathbb{R}_{t} < 2.15 \;$ 
& $\; \hat{p} \:\!<\:\! 54 \:\!\mbox{\small \%} \;$ \\
\hline
\hline
\end{tabular}
\nl
}
\nl
\mbox{} \vspace{-0.200cm} \\
\mbox{} \hspace{+0.350cm} 
{\small {\bf Table\;2:}\!\;\!
\begin{minipage}[t]{12.00cm}
{\small
Estimates of herd immunity levels 
$ \hat{p} = 1 - 1/\mathbb{R}_t $
for {\footnotesize C}ovid-19 
in ten countries
in the period of 2020
following the \!75th day after
the application of the first 
intervention mesures.
Case I corresponds to
average latency and transmission 
periods of 3 and 10 days,
respectively,
while Case II assumes
longer periods of
5 and 14 days. 
All the computations 
are based on covid data
for each country
available at
\href{https://www.worldometers.info/coronavirus/}%
{{\color{blue} worldometers/coronavirus}}.
}
\end{minipage}
}
\mbox{} \vspace{+0.400cm} \\
%
%
%
%

{\bf 4. Closing Remarks} \\
\mbox{} \vspace{-0.575cm} \\

Estimates of herd immunity levels
are dependent on the underlying mathematical model 
and its assumptions and limitations, 
data quality and reliability, 
limitations of the mathematical methods employed, and   
various factors that may be difficult to predict
like changes in the pathogenic agent
or in the population behavior. 
Thus,
any results should be viewed with caution
and interpreted as guidelines 
and not as definitive answers. 
Of course, 
the estimates obtained here
are in no way different. 
Constant checking and updating
in face of new experimental evidence
is necessary. 

For example, 
we have assumed that underreporting levels
of new covid cases in each surveyed country 
remained essentially constant along 2020,
so that their effects
on reproduction numbers 
are negligible
\cite{Cori2013}. 
But this may not have been quite so. 
In fact, 
data quality is commonly a major difficulty
in the study of an epidemic.
Still,
it seems safe to say
that herd immunity for Covid-19
could be close to 
90\:\!\mbox{\small \%},
at least for
sufficiently homogeneous populations.
Good hygienic practices 
and mask wearing
can very likely reduce
these values to 70\:\!\mbox{\small \%}
or lower,
depending on the case. 
As transmission rates 
may vary significantly
across regions,
even in the same country, 
local analysis is advised.
Finally,
our estimates provide basic upperbounds
for more realistic models
and similar studies
considering heterogeneous populations.


%
%

%
%

\end{document}